\documentclass[aps,twocolumn,floats,prd,nofootinbib,10pt]{revtex4-1}
\usepackage{amsmath}
\usepackage{amssymb}
\usepackage{amsthm}
\usepackage{graphicx,graphics,epsfig}
\usepackage{wrapfig}
\usepackage{natbib}
\usepackage{enumerate}

\def\VEV#1{{\left\langle #1 \right \rangle}}

\begin{document}

\title{Circular polarization in a spherical basis}

\author{Marc Kamionkowski}

\affiliation{Department of Physics and Astronomy, Johns Hopkins
	University, 3400 N.\ Charles Street, Baltimore, MD 21218, USA}

\begin{abstract}
Circular polarization of the cosmic microwave background (CMB)
arises in the standard cosmological model from Faraday
conversion of the linear polarization generated at the
surface of last scatter by various sources of birefringence
along the line of sight.  If the sources of birefringence
are generated at linear order in primordial density perturbations 
the principal axes of the index-of-refraction tensor are
determined by gradients of the primordial density field.  Since
linear polarization at the surface of last scatter is generated
at linear order in density perturbations, the
circular polarization thus arises at second order in primordial
perturbations.  Here, we re-visit the calculation of the
circular polarization using the total-angular-momentum formalism,
which allows for some simplifications in the calculation of the
angular power spectrum of the circular polarization---especially
for the dominant photon-photon scattering contribution---and
also provides some new intuition.
\end{abstract}
\maketitle

\section{Introduction}
\label{sec:intro}

The polarization of the cosmic microwave background (CMB) is the
subject of considerable study, as its detailed characterization
can constrain cosmological parameters and the mass distribution at
intermediate and low redshifts, and perhaps lead to insights about
the early Universe \cite{Kamionkowski:1999qc,Kamionkowski:2015yta,Abazajian:2016yjj}.
The polarization that is being sought is generally restricted,
though, to the linear polarization that is generated at linear order
in the amplitude of primordial perturbations.

Circular polarization is generated by Faraday conversion of the
primordial linear polarization as CMB photon propagate through a
birefringent medium along the line of sight \cite{Cooray:2002nm}.
There are several ways that birefringence can arise at linear
order in the primordial density perturbation; for example, spin
polarization of neutral hydrogen atoms during the cosmic dark ages
\cite{Montero-Camacho:2018vgs} and light-light scattering
induced by quantum-electrodynamic effects
\cite{Motie:2011az,Sawyer:2014maa,Ejlli:2016avx,Sadegh:2017rnr,Montero-Camacho:2018vgs}---the
latter is expected to be the dominant effect.
Thus, circular polarization arises at second order in
primordial perturbations.  This circular polarization has been
derived in considerable detail in recent work, especially
that of Ref.~\cite{Montero-Camacho:2018vgs}, which provided the principal
inspiration for this work.

Here we revisit the calculation of fluctuations in circular
polarization using the total-angular-momentum formalism
\cite{Dai:2012bc,Dai:2012ma}.  This formalism was
developed to facilitate the predictions of cosmological models
for observables on the celestial sphere.  The calculation of the
circular polarization provides a nice example of
the power of the formalism.  While no new physics is presented
in this paper, there may be calculational simplifications and
physical insights presented by the approach.  Generalization to
exotic early-Universe sources of circular polarization (e.g.,
from birefringence associated with primordial magnetic fields
\cite{Bavarsad:2009hm}, neutrinos \cite{Mohammadi:2013dea} or
primordial gravitational waves) may also be simplified with this
approach.  The utility will be less apparent for late-Universe
effects like those involving the first stars
\cite{De:2014qza,King:2016exc}.

Current upper limits to the circular-polarization power spectrum
are $l(l+1) C_l^{VV} \lesssim (200\, \mu{\rm K})^2$ over the
multipole range $33 <l<307$ \cite{Nagy:2017csq} and $\lesssim
(1000\, \mu{\rm K})^2$ at larger angular scales
\cite{Mainini:2013mja}.  These are orders of magnitude
above the expected primordial signal.
Still, renewed attention on the
circular polarization is warranted given the prospects with
CLASS \cite{Essinger-Hileman:2014pja} for improvement in the
sensitivity, by several orders of magnitude, to CMB circular
polarization and thus for the prospects of hitherto
unanticipated signals from new physics.

Below, we begin with a brief review of the
total-angular-momentum (TAM) formalism, focussing specifically
on the ingredients relevant for the circular-polarization
calculation.  We then show how the circular-polarization power
spectrum is obtained in the TAM formalism, using relevant
physics results from prior work.  We first discuss the power
spectrum from photon-photon scattering, which provides the
dominant effect.  We then discuss the power spectrum from a
subdominant contribution due to spin polarization in hydrogen
atoms, as this illustrates other aspects of the TAM approach.

\section{Total angular momentum waves}
\label{sec:TAM}

The plane waves $e^{i \vec k\cdot \vec x}$ constitute a simple
and familiar complete orthonormal basis for
scalar functions in Euclidean 3-space.  They are easily
generalized to vector fields with the adoption of three
polarization vectors $\epsilon_i(\vec k)$ for each Fourier mode
and likewise to symmetric trace-free tensor fields with
symmetric trace-free polarization tensors $\epsilon_{ij}(\vec
k)$.  The formalism becomes unwieldy for cosmological
observations, though, when projecting these plane waves onto the
spherical surface of the sky.

The total-angular-momentum (TAM) formalism
\cite{Dai:2012bc,Dai:2012ma} provides an alternative set of
basis functions, for scalar, vector, and tensor fields in
Euclidean 3-space, that are easily amemable to
the calculation of observables on the celestial sphere.  As we
will see, TAM waves also provides geometric insights in some
cases.  We begin with a scalar function $\phi(\vec x)$ which can
be expanded as
\begin{equation}
     \phi(\vec x) = \sum_{klm} \phi^k_{lm} \Psi^k_{lm}(\vec x),
\label{eqn:scalarTAM}     
\end{equation}
in terms of scalar TAM waves $\Psi^k_{lm}(\vec x)$.  Here $k$
is a wavenumber, $\sum_k$ is a shorthand for $\int k^2\,
dk/(2\pi)^3$, and the scalar TAM wave is $\Psi^k_{lm} = 4 \pi i^l
j_l(kx) Y_{lm}(\hat x)$ in terms of spherical Bessel functions
and spherical harmonics.\footnote{Our definition of
the harmonics differ by a factor of $4\pi i^l$ from that in
Ref.~\protect\cite{Dai:2012bc}, a choice that makes for more
compact expression.}  The orthonormality properties of these
functions imply that the scalar TAM-wave expansion coefficients
are
\begin{equation}
     \phi^k_{lm} = \int d^3x\, \phi(\vec x)
     \left[\Psi^k_{lm}(\vec x) \right]^*.
\end{equation}
If $\phi(\vec x)$ is a realization of a statistically isotropic
and homogeneous random field, then
\begin{equation}
     \VEV{ \phi^k_{lm} \left(\phi^{k'}_{l'm'} \right)^*} =
     \delta_{kk'} \delta_{ll'}\delta_{mm'} P_\phi(k),
\end{equation}
where $P_\phi(k)$ is the power spectrum for $\phi$, and
$\delta_{kk'}$ is shorthand for $(2\pi)^3 k^{-2}
\delta_D(k-k')$, with $\delta_D(k-k')$ a Dirac delta function.

Now consider a symmetric ($T_{ij}=T_{ji}$) trace-free
($T^i_{\,\, i}=0$) tensor $T_{ij}(\vec x)$, which involves
five degrees of freedom most generally.\footnote{We use
indexes $ij\ldots$ for three-dimensional tensors and following
Ref.~\protect\cite{Kamionkowski:1996ks} use 
indexes $ab\ldots$ for their projections onto the
two-dimensional celestial sphere.}  This can be expanded in
terms of five sets of tensor TAM waves
$\Psi^{\alpha,k}_{(lm)ij}(\vec x)$, for $\alpha=$L,VE,VB,TE,TB.
Here, L refers to the longitudinal mode, and VE/VB and
TE/TB to vector and tensor modes, respectively.  Of relevance
here is that the longitudinal mode has no variation in the plane
orthogonal to the direction of the gradient of $T_{ij}$, while
the other four modes do.  Given that all the physical effects we
will deal with in this paper trace back to primordial scalar
perturbations, which can have no variation in the plane
orthogonal to the direction of the gradient, we will need in
this paper only the longitudinal TAM wave.  This is obtained
from the scalar TAM wave from [Eq.~(81) in 
Ref.~\cite{Dai:2012bc}],
\begin{equation}
     -\nabla^2 \Psi^{{\rm L},k}_{(lm)ij}(\vec x) = \sqrt{\frac{3}{2}} \left(
     \nabla_i \nabla_j - \frac{1}{3} g_{ij} \nabla^2 \right)
     \Psi_{lm} (\vec x),
\label{eqn:scalartensor}     
\end{equation}
where $\nabla$ is a derivative with respect to $\vec x$ and
$g_{ij}$ a metric in Euclidean 3-space.

As we will see shortly, the circular polarization and
index-of-refraction tensors are projections of a
three-dimensional tensor onto the two-dimensional
celestial sphere.  That is, the polarization tensor in a direction
$\hat n$ will be
$P_{ab}(\hat n) \propto {\cal P}_a^{\, i}(\hat n) {\cal P}_b^{\, j}(\hat n)
T_{ij}(\hat n \chi)$, where ${\cal P}^{ab}(\hat n) = g^{ab} -\hat n^a
\hat n^b$ projects onto the plane orthogonal to $\hat n$.  Since
we deal here only with mechanisms that involve primordial
density perturbations, we will deal only with longitudinal
tensors.  The final key TAM-wave
result we need then comes from Eq.~(94) in Ref.~\cite{Dai:2012bc},
which provides the projection of the longitudinal-tensor TAM
wave onto the celestial sphere.  The result is,
\begin{equation}
     {\cal P}_a^{\, i}(\hat n) {\cal P}_b^{\, j}(\hat n)
     \Psi^{{\rm L},k}_{(lm)ij}(\vec
     x) = -\frac{4 \pi i^l}{N_l} \sqrt{\frac32} \frac{
     j_l(kr)}{(kr)^2} Y^{\rm E}_{(lm)ab}(\hat n),
\label{eqn:projection}     
\end{equation}
where $Y^{\rm E}_{(lm)ab}(\hat n)$ is the E-mode tensor spherical
harmonic \cite{Kamionkowski:1996ks,Kamionkowski:1996zd},
normalized as in Ref.~\cite{Kamionkowski:2015yta} to correspond
to the conventions of
Refs.~\cite{Seljak:1996gy,Zaldarriaga:1996xe}, and $N_l \equiv
\sqrt{2 (l-2)! /(l+2)!}$.
Eq.~(\ref{eqn:projection}) is the central TAM-wave result we
will use.  Eq.~(\ref{eqn:projection}) also shows that the
absence of B modes from density perturbations
\cite{Kamionkowski:1996zd,Seljak:1996gy} is related to the
fact that the projection of a longitudinal-tensor field onto the
sky has no B mode.

\section{Calculation}

The circular polarization induced in a photon observed from
direction $\hat n$ with linear polarization described by Stokes
parameter $Q$ and $U$ is
\begin{equation}
     V(\hat n) = \phi_Q(\hat n) U(\hat n) -\phi_U(\hat n) Q(\hat
     n),
\end{equation}
where
\begin{equation}
     \phi_{Q,U}(\hat n) = \frac{2}{c} \int\, dr\, \omega(r)
     n_{Q,U}(\hat n r),
\end{equation}
are phase shifts for the $Q$ and $U$ polarizations.
Here, $\omega$ is the photon angular frequency, $n_{Q,U}(r)$ are
the indexes of refraction for the $Q$ and $U$ polarizations at a
distance $r$ (related to the comoving distance $\chi$ by $dr= a
d\chi$, where $a$ is the scale factor.
If we take $\hat n$ in the $\hat z$ direction, the
index-of-refraction tensors are $n_Q = (1/2)(n_{xx}-n_{yy})$ and
$n_U= n_{xy}$, where $n_{ij} \propto \chi_{ij}\propto \partial_i
\mu_j$ are components of the Cartesian index-of-refraction
tensor and $\chi_{ij}$ is the magnetic-susceptibility tensor.

The Stokes parameters are a spin-2 field under rotations about
the $\hat n$ axis.  Equivalently they can be represented as the
components of a symmetric trace-free tensor,
\begin{equation}
     P_{ab}(\hat n) = \frac{1}{\sqrt{2}}\left( \begin{matrix}
     Q(\hat n) & U(\hat n) \\ U(\hat n) &
     -Q(\hat n) \end{matrix} \right).
\end{equation}
Likewise, the phase shifts are components of a tensor,
\begin{equation}
     \Phi_{ab}(\hat n) = \frac{1}{\sqrt{2}}\left(\begin{matrix}
     \phi_Q(\hat n) &
     \phi_U(\hat n) \\ \phi_U(\hat n) &
     -\phi_Q (\hat n) \end{matrix} \right).
\label{eqn:phitensor}     
\end{equation}
From these two tensors we can construct two scalars: a tensor
analog $P^{ab} \Phi_{ab}$ of a dot product, and a tensor analog
of a cross product, which we identify as the circular
polarization,
\begin{equation}
     V(\hat n)= \epsilon_{ac} P^{ab}(\hat n) \Phi_{b}^{\ c}(\hat
     n),
\label{eqn:V}     
\end{equation}
where $\epsilon_{ac}$ is the antisymmetric tensor on the
celestial sphere (see, e.g., Ref.~\cite{Kamionkowski:1996ks}).

If we consider arbitrary photon trajectories, the
propagation of the different linear-polarization states will be
described by a three-dimensional cartesian
tensor $n_{ij}$, which we can take to be traceless $n^i_{\, i}=0$,
since the relevant observables in this paper will depend only on
phase differences.  If birefringence is generated by the
primordial density field, the index-of-refraction tensor
$n_{ij}(\vec x)$ must be related to the density field by
\begin{equation}
     n_{ij}(\vec x) \propto \left(\nabla_i \nabla_j -
     \frac{1}{3} g_{ij} \nabla^2 \right) \delta(\vec x).
\end{equation}
This is the only STF tensor field that can be constructed from
$\delta(\vec x)$ that has no variation in the two directions
orthogonal to $\vec \nabla \delta(\vec x)$.  

Now we return to the calculation of the circular-polarization
power spectrum.  This involves a linear-polarization field
provided in terms of $Q(\hat n)$ and $U(\hat n)$ as a function
of position $\hat n$ on the spherical sky.  These are components
of a spin-2 field and can thus be replaced by the geometrical
invariants $E(\hat n)$ and $B(\hat n)$.  This
linear-polarization field is presumed
to arise at linear order in primordial perturbations, and the
polarization generated at position $\chi\hat n$ at a comoving distance $\chi$
in the direction $\hat n$ is related to the components of the
quadrupole moment, orthogonal to $\hat n$, of the radiation
field at that point.  That is, the contribution to the polarization
tensor at position $\chi \hat n$ is  $P_{ab}(\hat n) \propto
({\cal P}\, {\cal P} \,n)_{ab}(\hat n)$, as discussed in Section~\ref{sec:TAM}.
In words, the observed
polarization field is related to the components of a
longitudinal-tensor field constructed from the density field.

The same arguments apply also to $\Phi_{ab}$ and $\phi_{Q,U}$.
To be precise (as it will be required in Section
\ref{sec:spinpolarization}), the relation between the
phase-shift tensor, on the celestial sphere, in terms of the
projection $({\cal P}{\cal P}n)_{ab}(\hat n)$ is
\begin{eqnarray}
     \Phi_{ab}(\hat n)  &=&\frac{\omega_0}{\sqrt{2}} \frac2c \int\, d\chi\,
     \left[ ({\cal P}{\cal P}n)_{ab}(\hat n \chi) \right. \nonumber \\
     & &  \left. \, \,\,  - \frac12
     ({\cal P}{\cal P}n)_c^{\, c}(\hat n \chi) g_{ab} \right],
\label{eqn:phitensorfull}     
\end{eqnarray}
where $\omega_0$ is the angular frequency today, and here
$g_{ab}$ is a metric on the 2-sphere.  The trace in the second
term will be irrelevant when we project in the next paragraph
onto the traceless tensor spherical harmonics.

As discussed in Section \ref{sec:TAM}, the
linear-polarization pattern generated by primordial density
perturbations can be expanded,
\begin{equation}
     P_{ab}(\hat n) =\sum_{lm} P_{lm} Y^{\rm E}_{(lm)ab}(\hat n),
\end{equation}
in terms of the E-mode tensor spherical harmonics.
The same arguments apply, however, also to the phase
shifts---only $\phi_E$ is non-zero, and $\phi_B=0$---and
likewise to $n_{E,B}(r \hat n)$.  We thus conclude that we can
expand the phase-shift tensor as
\begin{equation}
     \Phi_{ab}(\hat n) =\sum_{lm} \Phi_{lm} Y^{\rm E}_{(lm)ab}(\hat n).
\end{equation}

The spherical-harmonic coefficients of the circular
polarization, given by Eq.~(\ref{eqn:V}), are then
\begin{equation}
     V(\hat n) = \epsilon_{ab} P^{ac} \Phi^b_{\ c}.
\end{equation}     
The spherical-harmonic coefficients for the circular
polarization are then
\begin{eqnarray}
     V_{lm} &=& \sum_{l_1 m_1}\sum_{l_2 m_2} P_{l_1m_1}
     \Phi_{l_2 m_2} \nonumber \\
     &\times &\int d\hat n\, \epsilon^{ab} Y^{\rm E}_{(l_1
     m_1)ac}(\hat n) Y^{\rm E}_{(l_2
     m_2)b}{}^{\ c}(\hat n) Y^*_{lm}(\hat n).   
\end{eqnarray}
The factor $\epsilon^{ab} Y^{\rm E}_{(l_1 m_1)ac}(\hat
n)$ in the integrand is a tensor with indices $b$ and $c$ that
are then contracted with another tensor that is symmetric under
$b \leftrightarrow c$.  We can therefore replace $\epsilon^{ab}
Y^{\rm E}_{(l_1 m_1)ac}(\hat
n)$ with its symmetrized version, which is $Y^{\rm B}_{(l_1
m_1)}{}^b_{\ c}(\hat
n)$.  The integral over the product of the three spherical
harmonics is thus, with the help of Eq.~(21) in
Ref.~\cite{Gluscevic:2009mm}, evaluated to be
\begin{equation}
     G^{lm}_{l_1m_1l_2 m_2} = -\xi^{lm}_{l_1,-m_1,l_2m_2}
     H^l_{l_1l_2},
\end{equation}
with $\xi^{lm}_{l_1m_1l_2m_2}$ and $H^l_{l_1l_2}$ as defined in
terms of Wigner-3j symbols in 
Ref.~\cite{Gluscevic:2009mm}.  Note that $G^{lm}_{l_1m_1l_2
m_2}$ is nonzero only for $l+l_1+l_2$=odd, and also that
$G^{lm}_{l_1 m_1 l_2 m_2} = - G^{lm}_{l_2 m_2 l_1 m_1}$.  
We then find
\begin{equation}
     V_{lm} = \sum_{l_1 l_2 m_1 m_2} P_{l_1
     m_1} \Phi_{l_2 m_2} G^{lm}_{l_1 m_1 l_2 m_2}.
\end{equation}

We now calculate the angular power spectrum $C_l^{VV}$ for the
circular polarization.
Using the antisymmetry of $G^{lm}_{l_1m_1l_2m_2}$ under $l_1m_1
\leftrightarrow l_2m_2$, the circular-polarization power
spectrum is then,
\begin{equation}
     C_l^{VV} = \sum_{l_1 m_1 l_2 m_2} \left( C_{l_1}^{PP}
     C_{l_2}^{\Phi\Phi} - C_{l_1}^{P\Phi} C_{l_2}^{P\Phi}
     \right) \left| G^{lm}_{l_1 m_1 l_2 m_2}
     \right|^2.
\end{equation}
The minus sign arises because the circular polarization is a
cross-product between the two fields; if there is a
cross-correlation between $P$ and $\Phi$, it reduces the
contribution to $C_l^{VV}$ from what it would be if they were
uncorrelated.  The sum over $m_1,m_2$ can be can be done using
results from Ref.~\cite{Gluscevic:2009mm}, from which follows,
\begin{equation}
     C_l^{VV} = \sum_{l_1 l_2} \frac{(2l_1+1)(2 l_2+1)}{4\pi}
     \left[C_{l_1}^{PP} C_{l_2}^{\Phi \Phi} -
     C_{l_1}^{P\Phi} C_{l_2}^{P\Phi} \right] \left( H^l_{l_1 l_2}
     \right)^2,
\end{equation}
where $C_l^{PP}$ and $C_l^{\Phi\Phi}$ are the polarization and
$\Phi$ power spectra, respectively.  Here, $C_l^{P\Phi}$ is the
power spectrum for the cross-correlation between $P_{ab}$ and
$\Phi_{ab}$; i.e., 
\begin{equation}
     \VEV{P_{l m} \Phi_{l'm'}^*} = C_l^{P\Phi}
     \delta_{ll'} \delta_{mm'}.
\end{equation}     

The flat-sky limit of this expression can be obtained in the
limit $l,l_1,l_2 \gg 1$ following the discussion in Section IV
in Ref.~\cite{Gluscevic:2009mm},
\begin{equation}
     C_l^{VV} = \int \frac{d^2l_1}{(2\pi)^2} \sin^2 2
     \varphi_{\vec l_1,\vec l -\vec l_1} \left(C_{l_1}^{PP}
     C_{|\vec l-\vec l_1|}^{\Phi\Phi} - C_{l_1}^{P\Phi}
     C_{\vec l -\vec l_1|}^{P\Phi} \right),
\label{eqn:Vpowerspectrum}     
\end{equation}
and the rms circular polarization is,
\begin{equation}
     \VEV{V^2} = \int \frac{d^2 l}{(2\pi)^2} C_l^{VV} = \frac12
     \left(\VEV{P^2} \VEV{\Phi^2} - \VEV{P \Phi}^2 \right).
\end{equation}

To summarize, the principal result of this Section is the
representation of  $C_l^{VV}$ in terms of the power spectra for
$P$ and $\Phi$ and their cross-correlation.

\subsection{Photon-photon scattering}

All that remains to do is to calculate the power spectra, but
the hard atomic-physics/QED work has been done for us already
\cite{Montero-Camacho:2018vgs}.  The advantage we will obtain
here is from the calculation of the angular fluctuations.  We
treat first the dominant contribution, from photon-photon
scattering, to the birefringence.

The
primordial curvature perturbation $\zeta(\vec x)$ is
expanded as in Eq.~(\ref{eqn:scalarTAM}) in terms of curvature
TAM coefficients $\zeta^k_{lm}$.  Any observable on the
celestial sphere of quantum numbers $lm$ then receives
contributions only from $\zeta^k_{lm}$ of the same $lm$.
The TAM-wave
amplitudes are related to the more familiar Fourier amplitudes
$\zeta(\vec k)$ through $\zeta^k_{lm} = \int d\hat k \zeta(\vec
k)$, and the transfer function depends only on the
magnitude $k=|\vec k|$, not its orientation $\hat k$.
Therefore, the
transfer function that determines the contribution of each TAM
wave $\zeta^k_{lm}$ to the observable is the same for the
TAM wave as for the Fourier mode of the same $k$.  We can thus
write, using Eq.~(\ref{eqn:projection}), for $X=\{P,\Phi\}$,
\begin{equation}
     X_{lm} = 4\pi N_l \int d\chi \sum_k \zeta^k_{lm} T_X(k,\eta_0-\chi)
     \frac{j_l(k \chi)}{(k\chi)^2}.
\end{equation}
The integral here is along the line of sight, or equivalently,
in conformal time, and $T_X(k,\eta_0-\chi)$ is
a transfer function that determines the contribution from
conformal time $\eta_0-\chi$ (where $\eta_0$ is the conformal time
today) to the polarization or phase $\Phi$ from a TAM wave (or
Fourier mode), of wavenumber $k$, of the primordial curvature
perturbation $\zeta$.

The transfer function for the polarization is provided by
numerical Boltzmann codes and inferred, e.g., in the
line-of-sight formalism from Eq.~(17) in
Ref.~\cite{Zaldarriaga:1996xe}.  If we take the last-scattering
surface to be thin, then
\begin{equation}
     T_P(k,\eta_0-\chi) = \frac{3}{4} \delta_D(\chi-\chi_{\rm
     ls}) \Pi(k,\eta_0-\chi),
\end{equation}
where $\chi$ is a comoving distance along the line of sight, the
second argument of $\Pi$ is the conformal time at that comoving
distance, and $\delta_D(\chi-\chi_{\rm ls})$ the Dirac delta
function.  Here, $\Pi(k,\eta)$ is the polarization source
function, defined as in Ref.~\cite{Zaldarriaga:1996xe}
(this is $2 {\cal P}^{(0)}$ in CLASS; cf.~Ref.~\cite{Tram:2013ima}).
As the discussion around
Eqs.~(5.19)--(5.23) in Ref.~\cite{Montero-Camacho:2018vgs}
indicates, the transfer function for the rotation is
\begin{equation}
     T_\Phi(k,r) = \frac{3\bar A}{4} (1+z)^4 \Pi(k,r_0-r),
\end{equation}
where
\begin{equation}
     \bar A = 1.76 \times 10^{-38} \left( \frac{\nu_0}{100\,
     {\rm GHz}} \right)\,{\rm m}^{-1}.
\end{equation}
The polarization and phase-shift power spectra (for $X_1 X_2
=\{PP,\Phi\Phi,P\Phi\}$) are then
\begin{equation}
     C_l^{X_1 X_2} = (4\pi N_l)^2 \frac{9}{16} \int \frac{k^2\,dk}{(2\pi)^3}
     P_\zeta(k) F^{X_1}_l(k) F^{X_2}_l(k),
\end{equation}
where the polarization window function is
\begin{equation}
     F^P_l(k) = \Pi(k,\eta_{\rm ls}) \frac{j_l(k \chi_{\rm
     ls})}{(k \chi_{\rm ls})^2},
\end{equation}
and the phase-shift window function is
\begin{equation}
     F^\Phi_l(k) = \bar A \int\, d\chi\ (1+z)^4
     \Pi(k,\eta_0-\chi)\frac{j_l(k \chi)}{(k \chi)^2}.
\end{equation}     
The results for $C_l^{PP}$, $C_l^{\Phi\Phi}$, and $C_l^{P\Phi}$
are then plugged into Eq.~(\ref{eqn:Vpowerspectrum}) to obtain
$C_l^{VV}$.  The numerical results should agree with those shown
in Fig.~7 in Ref.~\cite{Montero-Camacho:2018vgs}.

We close by estimating the range of multipoles $l$ over which
the cancellation from cross-correlation of $P$ and $\Phi$ will
be significant.  For large $l$, the spherical Bessel functions
are approximated by $j_l(kx)=0$ for $kx<l$ and $j_l(kx) \sim
(1/kx)\sin(kx -l \pi/2)$ for $kx>l$.  There is thus a strong
contribution to the cross-correlation for wavenumbers $k
(\chi-\chi_{\rm ls}) \lesssim 1$, and the contribution to $C_l$ is
dominated by wavenumbers $k\sim l/\chi_{\rm ls}$.  Since the
phase-shift integrand is weighted by $(1+z)^4$, about half the
signal is contributed for redshifts $900 \lesssim z \lesssim
1100$ which corresponds to a comoving distance $\sim50$ Mpc.
Using $r_{\rm ls}\simeq14,000$ Mpc, we infer that the
cross-correlation will suppress the circular-polarization signal
that would otherwise arise, in the absence of the
cross-correlation, for multipoles $l\lesssim 300$.

\subsection{Spin polarization of neutral hydrogen}
\label{sec:spinpolarization}

We now consider the calculation of the phase-shift power
spectrum for the birefringence due to spin polarizations of
hydrogen atoms.  While this provides a subdominant effect, the
calculation will illustrate some additional aspects of the TAM
formalism.

Ref.~\cite{Montero-Camacho:2018vgs} derives, using a
spherical-tensor basis, a relation
between the $n_{xx} - n_{yy}$ components of the
index-of-refraction tensor (for propagation of a photon in the
$\hat z$ direction) from a single Fourier mode of the
density field with wavevector $\vec k$.  This relation is
\begin{equation}
     (n_{xx}-n_{yy})_{\vec k} = - \frac{c p W}{\omega_0} \delta(\vec k) \left[
     Y_{22}(\hat k) +Y_{2,-2}(\hat k) \right],
\end{equation}
where $p$ and $W$ are as defined in their paper.  Using
\begin{equation}
     Y_{22}(\hat k) + Y_{2,-2}(\hat k) = 2 \sqrt{\frac{15}{32\pi}}
     (k_x^2-k_y^2),
\end{equation}     
we infer that in configuration space, the symmetric traceless
index-of-refraction tensor is related to the matter perturbation
$\delta(\vec x)$ through,
\begin{equation}
     \nabla^2 n_{ij}(\vec x) = \sqrt{\frac{15}{8\pi}} \frac{c p W}{\omega_0}
     \left(\nabla_i \nabla_j - \frac{1}{3} \delta_{ij} \nabla^2
     \right) \delta(\vec x),
\end{equation}
with $p$ and $W$ defined as in
Ref.~\cite{Montero-Camacho:2018vgs}.  We expand
\begin{equation}
     \delta(\vec x) = \sum_{klm} \delta^k_{lm}
     \Psi^k_{lm}(\vec x),
\end{equation}
in terms of scalar TAM waves and
\begin{equation}
     n_{ij}(\vec x) = \sum_{klm} n^k_{lm}
     \Psi^{{\rm L},k}_{(lm)ij}(\vec x),
\end{equation}
in terms of longitudinal-tensor TAM waves.  Then, from
Eq.~(\ref{eqn:scalartensor}) we infer,
\begin{equation}
     n^k_{lm} = - \sqrt{\frac{5}{4\pi}} \frac{p
     W}{\omega_0} \delta^k_{lm}.
\label{eqn:ndelta}     
\end{equation}
The projection of $n_{ij}(\vec \chi)$, at a (comoving) position
$\vec \chi =\hat n \chi$, onto the plane of the sky is, using
Eq.~(\ref{eqn:projection}),
\begin{equation}
     ({\cal P}\, {\cal P}\, n)_{ab}(\hat n)=
     \sum_{klm} n^k_{lm} \frac{-4 \pi i^l}{N_l} \sqrt{\frac32}
     \frac{j_l(k\chi)}{( k\chi)^2} Y^{\rm E}_{(lm)ab}(\hat n).  
\label{eqn:nprojection}
\end{equation}
We then infer from
Eq.~(\ref{eqn:ndelta}) and Eq.~(\ref{eqn:nprojection}) that the
contribution of any given $klm$ TAM wave to the
spherical-harmonic coefficient $\Phi_{lm}$ is, using
Eq.~(\ref{eqn:phitensorfull}),
\begin{equation}
     \Phi^k_{lm} = 2 \sqrt{15 \pi} i^l
     \frac{p}{N_l} \int \, d\chi\,  W(\chi)
     \frac{j_l(k\chi)}{(k\chi)^2} \delta^k_{lm}.
\end{equation}
The angular power spectrum is then obtained by summing over all
$k$ for this given $lm$; i.e.,
\begin{eqnarray}
     C_l^{\Phi\Phi} &=& \frac{60\pi p^2}{N_l^2} \int
     \frac{k^2\, dk}{(2\pi)^3} P(k,\eta_0)\nonumber \\
     &\times & \left[ \int\, d\chi\,
     W(\chi) D(\eta_0-\chi) \frac{j_l(k \chi)}{(k\chi)^2} \right]^2,
\end{eqnarray}
where we have written the conformal-time $\eta$ dependence of the
matter power spectrum as $P(k,\eta)=P(k,\eta_0)\left[D(\eta) \right]^2$
in terms of the linear-theory growth factor $D(\eta)$.
For the broad line-of-sight distribution here, we can use the
Limber approximation \cite{Limber:1954zz,Rubin} [see, e.g., Eqs.~(2.20) and
(2.21) in Ref.~\cite{Breysse:2014uia}] together with $N_l \simeq
2l^{-4}$ for $l\gg 1$.  We then note that the Limber
approximation sets $k\chi =l$ to obtain
\begin{equation}
     C_l^{\Phi\Phi} = \frac{15}{8 \pi} p^2 \int
     \frac{c\, dz}{\left[\chi(z) \right]^2 H(z)}
     \left[W\left(\chi\left(z \right) \right)
     \right]^2 P\left(\frac{ l}{ \chi(z)},\eta-\chi \right). 
\end{equation}
This then agrees with Eq.~(3.22) in
Ref.~\cite{Montero-Camacho:2018vgs}, given that their
$C_l^{\phi_Q}$ is half of $C_l^{\Phi\Phi}$.\\

\section{Conclusion}

Here we have used the total-angular-momentum formalism to
re-derive results from prior work on the power spectrum of
circular polarization that arises, at second order in
cosmological perturbations, in the standard cosmological model.
This alternative derivation may be useful to provide a different
perpective on the result and possibly some additional intuition;
and it also (once the TAM formalism has been digested)
simplifies some of the calculation.  In some sense, the true
power of the TAM formalism is lost on problems that involve only
scalar perturbations---it is far more powerful for calculations
that involve vector and tensor perturbations.  Thus, for
example, it may prove of particular value in models for circular
polarization that involve magnetic fields and/or primordial
tensor perturbations.  Still, the relative simplicity of the
problem considered here may be valuable as a simple application
of the TAM formalism.

\begin{acknowledgments}

We thank P.\ Montero-Camacho and V.\ Poulin for useful comments
on an earlier draft.  This work was supported by NSF Grant
No.\ 1519353, NASA NNX17AK38G, and the Simons Foundation.

\end{acknowledgments}

\end{document}